# Free energy model of emotional valence in dual-process perceptions

Hideyoshi Yanagisawa [a] [*], Xiaoxiang Wu [a], Kazutaka Ueda [a], Takeo Kato [b]

[a] The University of Tokyo, 7-3-1 Bunkyo, Hongo, Tokyo 113-8656, Japan

[b] Keio University, 3-14-1 Hiyoshi, Kohoku-ku, Yokohama-shi, Kanagawa 223-8522, Japan

**\*Corresponding author**

Email: hide@mech.t.u-tokyo.ac.jp

**Author emails:**

*Xiaoxiang Wu:* wu-xiaoxiang292@g.ecc.u-tokyo.ac.jp

*Kazutaka Ueda:* ueda@hnl.t.u-tokyo.ac.jp

*Takeo Kato:* kato@mech.keio.ac.jp

## Abstract

An appropriate level of arousal induces positive emotions, and a high arousal potential may provoke negative emotions. To explain the effect of arousal on emotional valence, we propose a novel mathematical framework of arousal potential variations in the dual process of human cognition: automatic and controlled. A suitable mathematical formulation to explain the emotions in the dual process is still absent. Our model associates free energy with arousal potential and its variations to explain emotional valence. Decreasing and increasing free energy consequently induce positive and negative emotions, respectively. We formalize a transition from the automatic to the controlled process in the dual process as a change of Bayesian prior. Further, we model emotional valence using free energy increase (FI) when one tries changing one's Bayesian prior and its reduction (FR) when one succeeds in recognizing the same stimuli with a changed prior and define three emotions: "interest," "confusion," and "boredom" using the variations. The results of our mathematical analysis comparing various Gaussian model parameters reveals the following: 1) prediction error (PR) increases FR (representing "interest") when the first prior variance is greater than the second prior variance, 2) PR decreases FR when the first prior variance is less than the second prior variance, and 3) the distance between priors' means always increases FR. We also discuss the association of the outcomes with emotions in the controlled process. The proposed mathematical model provides a general framework for predicting and controlling emotional valence in the dual process that varies with viewpoint and stimuli, as well as for understanding the contradictions in the effects of arousal on the valence.

## 1. Introduction

Emotions consist of two dominant dimensions: arousal and valence (Lang, 1995; Russell, 1980), which is called *core affect*. Arousal and valence are correlated with neural activities in specific brain regions such as the orbitofrontal cortex and amygdala, respectively (Wilson-Mendenhall et al., 2013). However, the two dimensions are not independent, and arousal affects valence. Berlyne proposed a function of the hedonic response of arousal potential, suggesting that the appropriate level of arousal potential induces positive responses, whereas an extreme arousal level induces negative responses (Berlyne, 1970). Consequently, the hedonic responses form an inverse U-shaped function of arousal potential, termed the *Wudnt curve*. Arousal is caused by cognitive factors such as novelty, complexity, and conflict. Berlyne termed such factors *collative variables*. Experimental studies have shown that collative variables form the Wudnt curve in a variety of fields such as art (Marin et al., 2016; Miyamoto & Yanagisawa, 2021; Silvia, 2005), design (Hekkert et al., 2003; Hung & Chen, 2012), and food science (Giacalone et al., 2014).

Yanagisawa et al. formulated the emotion arousal caused by collative variables using information-theoretic quantities and the Wudnt curve function as a summation of reward and aversion functions of information quantities (Miyamoto & Yanagisawa, 2021; Ueda et al., 2021; Yanagisawa, 2021; Yanagisawa et al., 2019). Yanagisawa recently suggested that information-theoretic free energy represents a general form of arousal induced by collative variables (Yanagisawa, 2021).

Free energy has been recognized as a key quantity to unify brain theories (Friston, 2010). The principle of free energy minimization explains perception and action (Friston et al., 2006). In this context, free energy refers to uncertainty and the prediction error of signals in a Bayesian brain theory (Knill & Pouget, 2004). Free energy has recently been applied to explain emotional valence. Joffily and Coricelli formalized valence as the rate of change of free energy over time; they suggest that decreasing free energy (or uncertainty) induces positive emotions, whereas increasing it induces negative emotions (Joffily & Coricelli, 2013). Hesp et al. formalized valence as the variation of confidence (or precision) in one's action model based on (expected) free energy (Hesp et al., 2021). Confidence increases with positive emotions as the expected free energy decreases, and vice versa. Such formulations assume that emotional states reflect changes in the uncertainty represented by free energy over time (Clark et al., 2018; Seth & Friston, 2016; Wager et al., 2015). Emerging mathematical formulations provide deductive predictions of emotions and essential understandings of the mechanism based on the principle of the Bayesian brain, such as the free energy principle (Friston, 2010).

Dual-process theories suggest that human cognition consists of two systems described as implicit and explicit, or System 1 and System 2 (Evans, 2003). The implicit process is automatic, unconscious, rapid, high in capacity, and less effortful, whereas the explicit process is conscious, slow, limited in capacity, and effortful. Emotions have also been discussed in dual-process perspectives (Fiori, 2009; Graf & Landwehr, 2015; Gyurak et al., 2011). In a neuroscientific experiment on the dual system of emotions involved in judging facial preferences, it has been shown that rapid, automatic processing occurs in the nucleus accumbens (NAC), followed by slower, explicit processing of preference judgments by the orbitofrontal cortex (OFC) (Kim et al., 2007). The NAC and OFC activities occur in temporally separated phases, and the increased NAC activity transfers preference signals to the OFC, which determines preference.

In aesthetic psychology, the pleasure-interest model of aesthetic liking (PIA model) models emotions in the dual process using the concept of processing fluency (Graf & Landwehr, 2015). The theory of fluency in aesthetic liking proposes that the experience of processing fluency of stimuli directly feels good on an affective level (Reber et al., 2004). Similar effects can also be confirmed as the link between typicality and preferences and have been demonstrated in various visual stimuli: human faces (Langlois & Roggman, 1990; Rhodes & Tremewan, 1996), painting/patterns (Graf & Landwehr, 2017; Winkielman et al., 2006), and artifacts/natural entities (Halberstadt & Rhodes, 2000; Landwehr et al., 2013).

The PIA model suggests that the processing fluency of stimuli in the first automatic process (e.g., the first impression of an aesthetic object) induces emotions such as pleasure and displeasure, and the disfluency reduction in the second controlled process (e.g., understanding complex and novel object or the resolution of conflict) induces emotions such as interest and confusion. The positive effect of disfluency reduction has also been confirmed as the aesthetic "Aha" effect or impact of the perceptual insight in the field of psychological aesthetics (Chetverikov & Filippova, 2014; Muth et al., 2013). Such a fluency–disfluency paradigm explains the aesthetic preferences discrepancy between fluently processed objects, such as simple and typical objects, and difficult-to-process (disfluent) objects such as complex novel objects.

The idea of processing fluency in dual processes is also consistent with the theory that fast thinking is subject to "cognitive ease" and that individuals tend to think, choose, and act spontaneously according to associative principles that are easy to understand and process (Kahneman, 2011). It has also been shown that a network of multiple brain regions, called the default mode brain network, which includes the ventromedial prefrontal cortex and anterior and posterior cingulate cortex, is associated with the neural basis of fast thinking available in a state of high cognitive fluency (Vatansever et al., 2017).

Although such a fluency–disfluency model provides an explanation of emotions in the dual process, the mathematical formulation is largely undiscovered. A mathematical formulation can provide quantitative predictions of emotions under various conditions. It can help in designing and engineering sensory stimuli that evoke certain emotions such as interest in the second process of the dual process.

We propose a novel mathematical framework of emotions in the dual process by applying free energy dynamics to the fluency–disfluency paradigm. The framework adopts variational Bayesian inference by minimizing free energy to model perceptions of the dual process and uses free energy variations to represent emotions such as pleasure, interest, confusion, and boredom. We consider a reselection of a Bayesian prior by switching category recognition as the second explicit process and formalize the emotions of the second process using the free energy increase and its reduction in changing the Bayesian prior. Consequently, we apply the proposed framework to a Gaussian Bayesian model. With this model, we formulate the free energy increase and its reduction in the second process to represent the emotions and analyze the effect of parameters such as the distance between the priors' means, prediction errors from the priors, the priors' uncertainties, and likelihood variance on the emotions. We discuss the fundamental characteristics of perception and emotions in the dual process based on the proposed framework and the results of the Gaussian analysis.

## 2. Method

Fig. 1 shows the flow of our research methodology. We formulate perceptions of stimuli under a certain recognition category using a hierarchical Bayesian model and free energy minimization. Using the Bayesian model, we formulate a shift from the automatic to the controlled process in the dual process by switching the Bayesian prior to represent the second recognition category. In the dual-process shifting, we calculate three free energy variations: free energy reduction in the automatic process, and free energy increase and reduction in the controlled process. We associate these free energy variations with emotions such as pleasure, interest, confusion, and boredom based on the fluency–disfluency paradigm. Finally, we analyze the effect of Gaussian model parameters such as prediction errors (distance between means) and uncertainties (variances) on the extent of emotional valence to elucidate the condition of emotions.

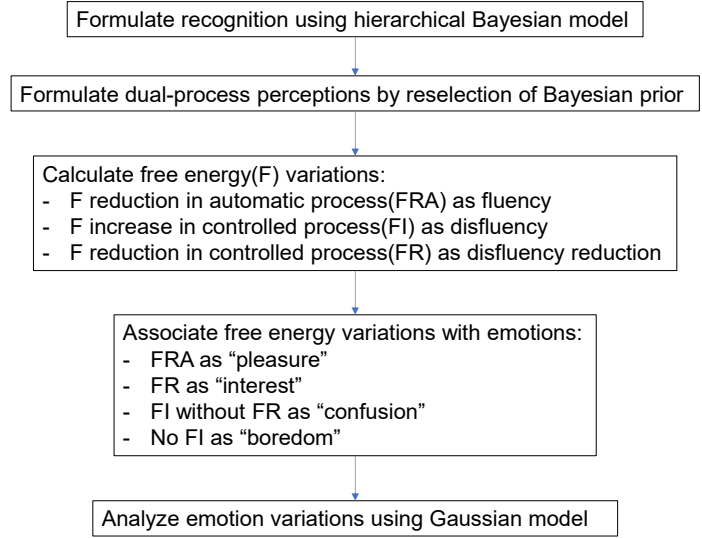

Fig. 1. Flow of our research methodology.

### 2.1 Bayesian perception and Free energy minimization

Bayesian brain theory suggests that brain activity is Bayes optimal (Knill & Pouget, 2004). Theoretical and empirical studies have supported the Bayesian brain theory from perception to motor-sensory optimizations (K. Friston, 2010; Kersten et al., 2004; Körding & Wolpert, 2004; Wei & Stocker, 2015; Yanagisawa, 2016). We adopt a Bayesian perception model in which that perception is an estimation of the cause of sensory outcomes, and these estimates follow Bayesian inference. The Bayesian theorem defines the posterior probability of cause $x$ given observation $y$ as proportional to the joint probability of $x$ and $y$.

$$p(x|y) = \frac{p(x,y)}{p(y)} \propto p(x,y) \qquad (1)$$

The distribution of joint probability can be assumed to be learned by experiencing varied combinations of $x$ and $y$ throughout one's life. We term the joint probability distributions *generative model* because it represents an internal model in the brain to predict $y$ under the estimation of $x$. The denominator of Eq. 1 is a marginal distribution, $p(y) = \int p(x,y)\,dx$, that represents evidence of the model when observation $y$ is obtained. The negative log of evidence, $-\log p(y)$, is termed Shannon's surprise meaning prediction error, which is the difference between the generative model's prediction and the encountering observation. The brain wants to estimate causes $x$ given $y$ to minimize the surprise (i.e., prediction error). The Shannon surprise is the lower bound of free energy. The free energy principle suggests that any self-organizing system (i.e., brain) that is at equilibrium with its environment must minimize its free energy (K. Friston, 2010; K. Friston et al., 2006). Free energy is an information quantity defined using the generative model $p(x,y)$ and *recognition density* $q(x)$ as shown in Eq.2.

$$F = \langle \log p(x,y) - \log q(x) \rangle_{q(x)} \tag{2}$$

With a decomposition $p(x,y) = p(x|y)p(y)$, free energy is formed as a summation of Kullback–Leibler divergence (KLD) from the true posterior to the recognition density and Shannon's surprise.

$$F = \langle \log q(x) - \log p(x|y) - \log p(y) \rangle_{q(x)} = D_{KL}[q(x)||p(x|y)] - \log p(y) \tag{3}$$

The KL-divergence, $D_{KL}[q(x)||p(x|y)] = \langle \log q(x) - \log p(x|y) \rangle_{q(x)}$, is always zero or above; thus, the free energy is the upper bound of Shannon's surprise, $F \geq -\log p(y)$. When the recognition density approximates the true posterior $q(x) \approx p(x|y)$, the KL-divergence reaches zero, and the free energy is minimized. This approximation process is called *perception*. In this sense, $q(x)$ is termed an approximate posterior meaning probability distribution of the perception. The free energy principle suggests that free energy is always reduced to Shannon's surprise in perception, and the percept approximately follows the Bayesian posterior.

With another decomposition $p(x,y) = p(y|x)p(x)$, free energy is represented as a summation of Bayesian surprise and inverse accuracy.

$$F = D_{KL}[q(x)||p(x)] + \langle -\log p(y|x) \rangle_{q(x)} \tag{4}$$

The Bayesian surprise, KL-divergence from the prior $p(x)$ to recognition density, represents information gain. It correlates with responses of human surprises and event-related potentials P300 (Yanagisawa et al., 2019). The accuracy denotes the extent to which the model prediction accurately matches the sensory observation. The inverse accuracy represents perceived uncertainty given sensory outcome $y$. It converges to represent perceived complexity after sufficient observations of the same sensory stimuli (Yanagisawa, 2021).

## 2.2 Category recognition and reselection of Bayesian prior

The same sensory stimuli may be recognized as multiple categories. A notable example is ambiguous images such as My Wife and My Mother-in-Law (WML) (Boring, 1942). One may initially recognize it as a young lady's picture (my wife), try to find an old lady (my mother), and then succeed in recognizing the old lady in the same picture by changing one's perspective. We associate such category switching from the young lady to the old lady with a transition from the automatic process to the controlled process of the dual process. Here, we formulate the category recognition transition of the dual process using a hierarchical Bayesian model. A set of recognition categories $C = \{c_1, ..., c_n\}$ and generative model of the $i$-th category are assumed as Eq. 5.

$$p_i(x,y) = p(x,y|c_i) = p(y|x)p(x|c_i) = p(y|x)p_i(x) \tag{5}$$

where $p(y|x)$ is an observation model of observation $y$ given feature $x$, and $p_i(x)$ is the prior of feature $x$ of the $i$-th recognition category $c_i$. Given observation $y$ under $c_i$, the free energy is formed as the summation of KL-divergence and surprise:

$$F_i := D_{KL}[q(x)||p_i(x|y)] - \log p_i(y) \tag{6}$$

where evidence $p_i(y) = \int_{-\infty}^{\infty} p_i(x,y)dx$, and true posterior $p_i(x|y) = \frac{p_i(x,y)}{p_i(y)}$ under the $i$-th category recognition. Using the free energy model, we formulate the free energy variations in the dual process. Figure 1 shows the schematic of free energy variation in the dual-process perceptions. When one recognizes $y$ as category $c_i$ in the first automatic process, the recognition density approximates the true posterior of the category, $q(x) \approx p_i(x|y)$, and the free energy is minimized to the surprise under the $i$-th category recognition.

$$F_{i,min} := -\log p_i(y) \tag{7}$$

This recognition process causes the free energy reduction as a KL-divergence from true posterior to approximate posterior, $D_{KL}[q(x)||p_i(x|y)]$. By applying Eq. 4, the minimized free energy is represented as the summation of Bayesian surprise and accuracy.

$$F_{i,min} = D_{KL}[p_i(x|y)||p_i(x)] + \langle -\log p(y|x) \rangle_{p_i(x|y)} \tag{8}$$

Now, when one tries to recognize the observation $y$ as another category $c_j$ ($j \neq i$) in the control process, the prior is replaced by $p_j(x)$. At that moment, the free energy is approximated in two forms as shown in Eq. 9. The first form is a summation of KL-divergence from the true posterior of category $c_j$ to the approximate posterior of $c_i$ and surprise based on $c_j$ prior. The second form is the summation of KL-divergence from the prior of $c_j$ to approximate posterior of $c_i$ and accuracy based on $c_j$ prior.

$$F_j := D_{KL}[p_i(x|y)||p_j(x|y)] - \log p_j(y) = D_{KL}[p_i(x|y)||p_j(x)] + \langle -\log p(y|x) \rangle_{p_i(x|y)} \tag{9}$$

Note that, the recognition density is still approximated to the true posterior of category $c_i$, $q(x) \approx p_i(x|y)$. If one succeeds to recognize $y$ as category $c_j$, the recognition density is approximated as the true posterior of $c_j$, $q(x) \approx p_j(x|y)$, and the free energy is minimized to surprise under category $c_j$. By applying Eq. 4, the surprise equals Bayesian surprise and inverse accuracy under-recognition of category $c_j$.

$$F_{j,min} := -\log p_j(y) = D_{KL}[p_j(x|y)||p_j(x)] + \langle -\log p(y|x) \rangle_{p_j(x|y)} \tag{10}$$

### 2.3 Free energy dynamics in dual process and fluency

In the dual process, there are three free energy variations: reduction in automatic process $\Delta F_i$, increased $\Delta F_{ij}$ and its reduction $\Delta F_j$ in a controlled process defined by Eqs. 11, 12, and 13, respectively (see also Fig. 2).

$$\Delta F_i := F_i - F_{i,min} \tag{11}$$

$$\Delta F_{ij} := F_j - F_{i,min} \tag{12}$$

$$\Delta F_j := F_j - F_{j,min} \tag{13}$$

$\Delta F_i$ is the free energy reduction by recognizing sensory stimuli as the category $c_i$ in the automatic process (e.g., recognition of my wife in WML). $\Delta F_{ij}$ represents the increase in free energy when one tries to recognize the same stimuli as another category $c_j$, but one does not recognize it yet. For example, in the WML case, one tries to find the mother in the picture but does not see it yet. $\Delta F_j$ is the second free energy reduction when one's controlled process succeeds to recognize the same stimuli as another category $cj$ (e.g., one finally finds the mother in WML). We discuss the association of these three free energy variations with the fluency paradigm of the PIA model in the following sections.

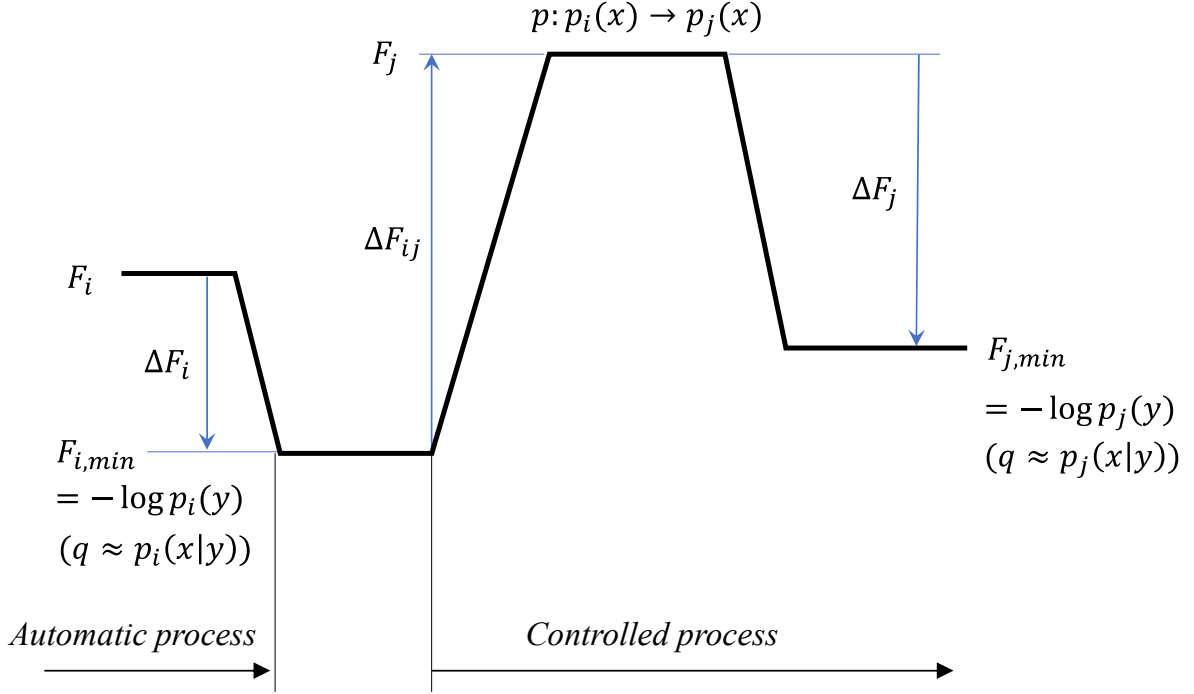

Fig. 2. Free energy variations in the dual process and emotions. The controlled process is formalized as a prior change and the recognition of the same stimuli with the changed prior.

### 2.4 Free energy reduction in the automatic process and fluency

By applying Eqs. 6 and 7 to Eq. 11, free energy reduction in the automatic process equals the KL-divergence from posterior to the initial approximate posterior.

$$\Delta F_i = D_{KL}[p_0(x)||p_i(x|y)] \geq 0 \qquad (14)$$

where $p_0(x)$ is an initial form of the approximate posterior, $q(x) \approx p_0(x)$. The KL-divergence represents the contents that one's brain gains and processes in the recognition process (Yanagisawa, 2021). The automatic process is unconscious, and one instantly experiences the recognition. In this case, the duration of the process in one's perception is infinitely small, and thus the free energy reduction represents the speed of information processing in the recognition, i.e., $dF/dt = \lim_{t \to 0} \Delta F_i/t$. The faster the information processing speed, the more fluent the information processing. Therefore, we consider that the free energy reduction represents the processing *fluency* discussed (Graf & Landwehr, 2015).

### 2.5 Free energy increase in controlled process and disfluency

By applying Eqs. 7 and 8 to Eq. 12, we find that the free energy increase in the controlled process comprises two KL divergences: KL-divergence from the prior of category $c_j$ in a controlled process to the true posterior of category $c_i$ in an automatic process, and the negative Bayesian surprise when recognized as category $c_i$ in an automatic process.

$$\Delta F_{ij} = D_{KL}[p_i(x|y)||p_j(x)] - D_{KL}[p_i(x|y)||p_i(x)] \qquad (15)$$

The former KL-divergence represents how far the prior of the second category $c_j$ is from the recognition of initial category $c_i$. We consider it difficult to switch the recognition from category $c_i$ to $c_j$ in the controlled process. KL-divergence is assumed to be greater than the Bayesian surprise of the automatic process because the recognition of $c_j$ in the controlled process is harder than $c_i$ in the automatic process. The greater the increase in free energy, the harder it is to recognize the same sensory stimulus as category $c_j$. This increase in free energy works as a "dam" of the information process flow. Therefore, we consider that the free energy increase corresponds to the concept of *disfluency* in the PIA model (Graf & Landwehr, 2015). This free energy increase seems to explain the increase in perceived complexity right before the perceptual insight in experimental research (Muth, Raab, et al., 2015).

The free energy increase is deformed to the difference between the two priors' information averaged over the posterior of the automatic process.

$$\Delta F_{ij} = \langle \log p_i(x) - \log p_j(x) \rangle_{p_i(x|y)} \tag{16}$$

Thus, the increase in free energy is greater when the difference between the two categories' priors is greater. This suggests that the farther the distance between the two priors, the harder it is to switch category recognition in the controlled process. Eq. 16 is equivalent to the difference of the two cross entropies:

$$\Delta F_{ij} = \langle -\log p_j(x) \rangle_{p_i(x|y)} - \langle -\log p_i(x) \rangle_{p_i(x|y)} \tag{17}$$

The first term on the righthand side of Eq. 17 denotes the deviation of the prior used in the controlled process from the true posterior of the leading automatic process, and the second term denotes the deviation of the prior used in the automatic process from the true posterior of the automatic process. We can assume that the deviation of the controlled process (the first term of Eq. 17) is greater than that of the automatic process (the second term of Eq. 17). Under this assumption, $\Delta F_{ij}$ is always positive (i.e., free energy always increases in the controlled process). Eq. 17 suggests that the increase in free energy is greater when the posterior in the automatic process is farther from the category recognized in the controlled process by comparing it with the distance from the category recognized in the automatic process.

## 2.6 Free energy reduction in controlled process and disfluency reduction

By applying Eqs. 9 and 10, to Eq. 13, the free energy reduction in the controlled process is equivalent to a KL-divergence from the posterior of the controlled process to the posterior of the automatic process.

$$\Delta F_j = D_{KL}[p_i(x|y)||p_j(x|y)] \geq 0 \tag{18}$$

Thus, the greater the difference in recognized features between automatic and controlled processes, the greater the free energy reduction. This free energy reduction implies process fluency in the controlled process by succeeding in the recognition switching. Therefore, we consider that the free energy reduction represents the concept of *disfluency reduction* in the PIA model (Graf & Landwehr, 2015).

From Eqs. 12 and 13, the free energy reduction is formed as the summation of the free energy increase of the controlled process and the decrease in minimized free energies of the two processes.

$$\Delta F_j = \Delta F_{ij} + \Delta F_{ij,min} \tag{19}$$

where $\Delta F_{ij,min} = F_{i,min} - F_{j,min}$. The first term suggests that the greater the increase in free energy at the controlled process, the greater the free energy reduction at the same process. In other words, the greater the disfluency, the greater the disfluency reduction. The second term represents the decrease in free energy between the two process recognitions. This suggests that the greater the difference in minimized free energy, the greater the free energy reduction.

## 2.7 Emotions and free energy variations

The free energy reduction in the automatic process $\Delta F_i$ induces positive emotions owing to its uncertainty reduction (Joffily & Coricelli, 2013). We associate it with process fluency. The PIA model suggests that fluency more than the expectation in the automatic process denotes "pleasure" (Graf & Landwehr, 2015). Therefore, we consider that free energy reduction induces positive emotions such as "pleasure."

The combination of the decrease and increase in free energy denotes positive and negative emotions, respectively. The PIA model suggests three cases of emotions, interest, confusion, and boredom, in the transition between automatic and controlled processes (Graf & Landwehr, 2015). The successful disfluency reduction in the controlled process induces "interest." If one fails to reduce the disfluency in the controlled process by solving, one perceives "confusion" as a negative emotion. If no disfluency occurs, one may feel "boredom." By adapting free energy increases $\Delta F_{ij}$ and decreases $\Delta F_j$ to disfluency and its reduction in the PIA model, respectively, we define the three emotions as the following conditions.

$$interest: \Delta F_{ij} > c_d \text{ and } \Delta F_j > c_r \quad (20)$$

$$confuse: \Delta F_{ij} > c_d \text{ and } \Delta F_j \leq c_r \quad (21)$$

$$boredom: \Delta F_{ij} \leq c_d \text{ and } \Delta F_j \leq c_r \quad (22)$$

where $c_d$ and $c_r$ are the thresholds of free energy increase and reduction, respectively. Figure 2 shows the schematic of the association of free energy deviations with emotions. "Interest" is defined as the occurrence of both a free energy increase (disfluency) and its reduction. We consider that a greater free energy reduction means more interest. The increase in free energy (disfluency) increases its reduction. Thus, the greater the free energy increase, the more interest. "Confusion" is defined as the case where free energy increase (disfluency) occurs but its reduction does not occur. "Boredom" is defined as the case where the increase in free energy (disfluency) does not occur; hence, its reduction does not occur.

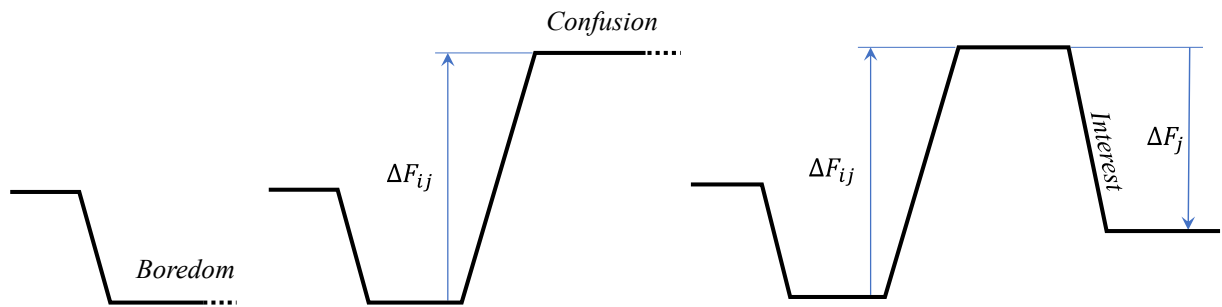

Fig. 3. Three emotions—boredom, confusion, and interest—as patterns of free energy variations. If one does not find any other perspectives to recognize the same stimuli differently, no free energy increase occurs; thus, one gets bored of the stimuli (i.e., no disfluency and its reduction) (left figure). If one does not succeed to recognize the second category, free energy increase (or disfluency) is not reduced, and one confuses the recognition (middle figure). If one succeeds in recognizing the second category, the free energy increase is reduced, and thus one is interested (right figure).

## 3. Results
### 3.1 Gaussian model of free energy in the dual process

The Gaussian form has been applied to analyze the characteristics of free energy in previous studies (Buckley et al., 2017; Yanagisawa, 2021). The Gaussian distribution consists of two independent parameters that represent expectation and uncertainty; specifically, mean and variance, respectively. The distance between the prior mean and likelihood peak is termed *prediction error* $\delta$, and the variance of prior variations is termed *prior uncertainty* $s_p$ in previous studies (Yanagisawa, 2021; Yanagisawa et al., 2019). The Gaussian form is useful to explicitly analyze the effects of the independent parameters of predictability on free energy.

The free energy of *n* observations randomly obtained from a stimulus source following a Gaussian distribution of variance $s_L$ is formed as a quadratic function of prediction error with coefficients of variances (for the derivation, see (Yanagisawa, 2021)):

$$F = A\delta^2 + B \qquad (23)$$

$$A = \frac{1}{2}\frac{n}{ns_p + s_L}, B = \frac{1}{2}\left\{\ln(ns_p + s_L) + (n-1)\ln s_L + n\ln 2\pi + n\frac{S}{s_L}\right\}$$

We consider the case of a single data observation *n*=1 to simplify further analysis. When *n*=1, the coefficients AF and BF are simplified as follows:

$$A = \frac{1}{2}\frac{1}{s_p + s_l}, B = \frac{1}{2}\{\ln(s_p + s_l) + \ln 2\pi\} \qquad (24)$$

The gradient $A$ is an inverse of variations representing the precision of information sources, that is, prior and stimuli. Intercept $B$ is a log variation, representing the uncertainty of the information sources. Therefore, free energy comprises a summation of prediction errors weighted by precision and uncertainty. This suggests that the higher the precision, the more sensitive the prediction error to free energy, and uncertainty increases free energy. In the dual process, we assume two Gaussian priors: prior of category *i* recognized in the automatic process: $p_i(x) = N(\mu_i, s_i)$ and prior of category *j* recognized in the controlled process: $p_j(x) = N(\mu_j, s_j)$. As shown in Fig. 4, we assume that the likelihood means is placed in-between these priors' means. Under this assumption, the difference between the priors' means $\mu_{ij}$ is a summation of the prediction errors for the two categories, $\delta_i$ and $\delta_j$. We analyze the effect of independent parameters such as prior distance $\mu_{ij}$, prediction error of category *j* in the controlled process $\delta_j$, and prior variances (uncertainties): $s_i$ and $s_j$ on three differences in free energy: 1) difference between minimized the free energy of the two categories to analyze the relative likelihood in category recognition, 2) free energy increase in the controlled process representing disfluency, and 3) free energy reduction in the controlled process representing disfluency reduction, in following sections.

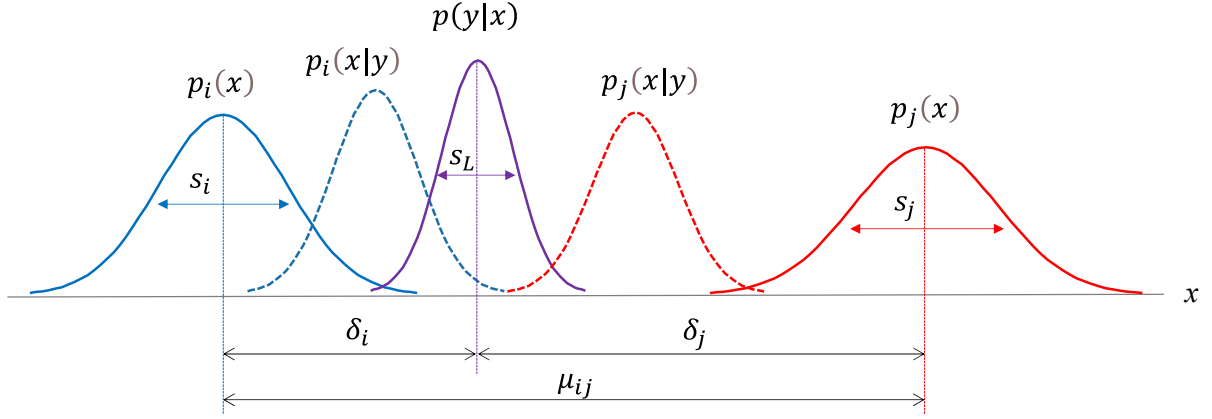

Fig. 4. Gaussian form of Bayesian priors and posteriors in the dual process. Category *i* is recognized in the automatic process with prior $p_i$, and then category *j* is recognized in the controlled process with prior $p_j$ from the same observation *y*. The likelihood of the observation is assumed to be placed in-between the two categories. The priors' mean distance $\mu_{ij}$ is a summation of the prediction errors $\delta_i$ and $\delta_j$. We use the ratio of the prediction error to the prior distance $r_j = \delta_j/\mu_{ij}, (0 \leq r_j \leq 1)$ as a relative prediction error in the controlled process.

### 3.2 Likelihood of recognition between two categories

Between the two categories recognized in the dual process, the less minimized the free energy, the more likely it will be recognized. We analyze the difference in minimized free energy between the two categories of the dual process, *i* and *j*, $\Delta F_{ij,min} = F_{i,min} - F_{j,min}$, using the Gaussian Bayes model (Eqs. 23 and 24). Specifically, we elucidate the effect of the relative prediction error $r_j$ and the priors' mean distance on the free energy difference. The free energy difference increases the free energy reduction $\Delta F_j$ as shown in Eq. 19. When the free energy difference is negative, $\Delta F_{ij,min} < 0$, category *i* is more likely to be recognized than category *j*. This condition is likely in the dual process because category *i* is easier to recognize than category *j* because the automatic process is effortless and the controlled process requires effort to recognize. However, the free energy difference can be positive, $\Delta F_{ij,min} > 0$, for the reason in the discussion.

The difference in minimized free energy is formed as a quadratic function of the prior distance by applying Eqs. 23 and 24.

$$\Delta F_{ij,min} = F_{i,min} - F_{j,min} = -\log p_i(y) + \log p_j(y)$$
$$= \frac{1}{2}\left\{\frac{\delta_i^2}{s_i + s_L} - \frac{\delta_j^2}{s_j + s_L} + \ln\frac{s_i + s_L}{s_j + s_L}\right\} \quad (25)$$

When the priors' variances are the same, i.e., $s_i = s_j = s_p$, the free energy difference is proportional to the difference in the square of prediction errors. This suggests that the category whose prior mean is closer to the likelihood peak is more likely to be recognized than the others.

$$\Delta F_{ij,min} = \frac{\delta_i^2 - \delta_j^2}{2(s_p + s_L)} \quad (26)$$

When the prediction errors are the same, $\delta_i^2 = \delta_j^2 = \delta^2$, i.e., the likelihood mean is equally distanced from the two priors' means, the free energy difference is formed as a quadratic function of the prediction error.

$$\Delta F_{ij,min} = \frac{1}{2}\left\{\frac{s_j - s_i}{(s_i + s_L)(s_j + s_L)}\delta^2 + \ln\frac{s_i + s_L}{s_j + s_L}\right\} = A\delta^2 + B \tag{27}$$

The distance between priors is $2\delta$. Thus, $\delta$ is proportional to the distance between priors. The coefficients A and B are functions of variances of the priors and likelihood. The signs of the coefficients are determined by the magnitude relations of the priors' variances.

$$\begin{aligned} A < 0, B > 0, & \quad (s_i > s_j) \\ A > 0, B < 0, & \quad (s_i < s_j) \\ A = 0, B = 0, & \quad (s_i = s_j) \end{aligned} \tag{28}$$

Figure 5 shows the difference in minimized free energy between categories as functions of relative prediction errors for different combinations of prior variations when the prediction errors of categories are the same. The free energy difference functions have an intersection with $\Delta F_{ij,min} = 0$, where $\delta^2 = \frac{(s_i+s_L)(s_j+s_L)}{s_j-s_i}\ln\frac{s_j+s_L}{s_i+s_L}$. The intersection is always positive or zero only when $s_i = s_j$. There are two cases where the minimized free energy of category $i$ is less than that of category $j$, $F_{i,min} < F_{j,min}$. If the prediction error (or the half of distance between priors) is zero or less than the intersection, a category is likely to be recognized if the prior variance is less than another. If the prediction error (or the half of distance between priors) is greater than the intersection, a category is likely to be recognized if the prior variance is greater than another. In addition, if the priors' variances are the same, the two categories are equally likely to be recognized regardless of the prediction error and the distance between priors.

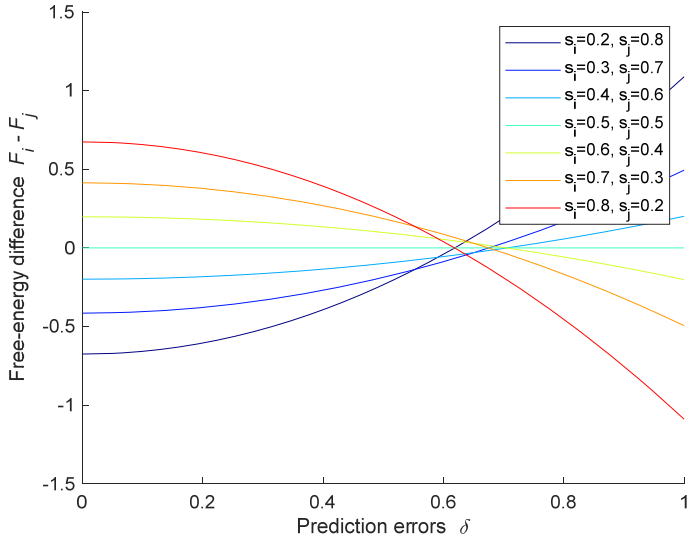

Fig. 5. Free energy difference representing relative likelihood as a function of the equal prediction errors for different prior variances.

By introducing the relative prediction error in the controlled process, $r_j = \delta_j/\mu_{ij}$ $(0 < r_j < 1)$, the minimized free energy difference $\Delta F_{ij,min}$ is represented as a quadratic function of the distance between priors' means $\mu_{ij}$.

$$\Delta F_{ij,min} = \frac{1}{2}\left\{\frac{(1-r_j)^2}{s_i + s_L} - \frac{r_j^2}{s_j + s_L}\right\}\mu_{ij}^2 + \frac{1}{2}\ln\frac{s_i + s_L}{s_j + s_L} \tag{29}$$

When the distance between the priors' means is zero, a category is likely to be recognized if the prior variance is smaller than another. From the gradient in Eq. 29, the distance between priors' means decreases the free energy difference if the prediction error of category $i$ weighted by precision is less than that of category $j$, $\frac{(1-r_j)^2}{s_i+s_L} < \frac{\delta_j^2}{s_j+s_l}$. In the opposite case, $\frac{(1-r_j)^2}{s_i+s_L} > \frac{\delta_j^2}{s_j+s_l}$, it increases the free energy difference.

Figure 6 shows the free energy differences between categories as functions of the relative prediction errors for different prior variance combinations in different levels of prior distance. When the distance between the priors' means $\mu_{ij}$ is constant, the free energy difference always decreases as the relative prediction error $r_j$ increases.

$$\frac{\partial \Delta F_{ij,min}}{\partial r_j} = \left(\frac{r_j - 1}{s_i + s_L} - \frac{r_j}{s_j + s_L}\right)\mu_{ij}^2 < 0 \tag{30}$$

Consider the sign of the intercepts of the functions shown in Fig. 5 (free energy difference functions of relative prediction error). When the relative prediction error is zero, $r_j = 0$, the intercept forms a quadratic function of the distance between priors.

$$\Delta F_{ij,min} = \frac{1}{2}\left(\frac{\mu_{ij}^2}{s_i + s_L} + \ln\frac{s_i + s_L}{s_j + s_L}\right) \tag{31}$$

When $\mu_{ij}$ is zero, the sing of Eq. 31 is determined only by the magnitude relation of the prior variances. Namely, $F_{i,min} < F_{j,min}$ if $s_i < s_j$, and vice versa (i.e., $F_{i,min} > F_{j,min}$ if $s_i > s_j$). Therefore, the intercept is negative when $s_i < s_j$ and the difference between priors is small, i.e., $\mu_{ij}^2 < (s_i + s_L)\ln\frac{s_j+s_L}{s_i+s_L}$.

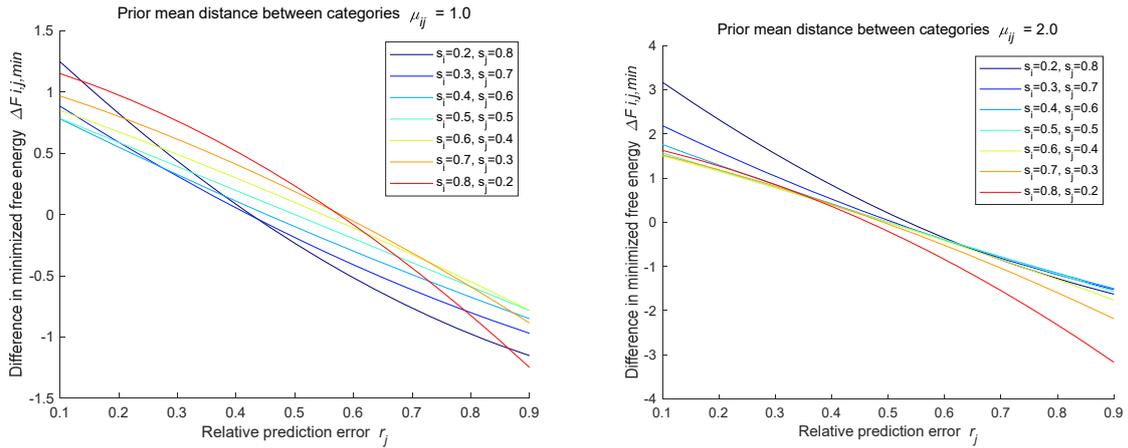

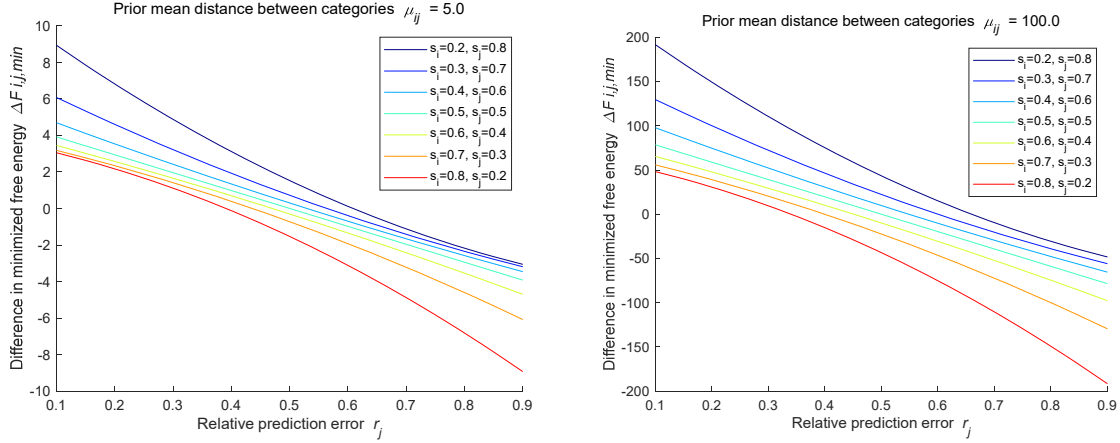

Fig. 6. Free energy difference as a function of relative prediction errors for different prior variances. The free energy difference decreases as the relative prediction error increases in all conditions. The function shape is concave when $s_i < s_j$, convex when $s_i > s_j$, and linear when $s_i = s_j$. The intercept increases as $s_j$ decreases when the distance between priors is sufficiently large ($\mu_{ij} = 5, 100$). However, when the distance between priors is small ($\mu_{ij} = 1.0, 2.0$), $s_i$ increases the intercept when $s_i > s_j$.

### 3.3 Disfluency: Free energy increase in the controlled process

In Section 2.7, we associated the free energy increase in the controlled process with the disfluency of the PIA model that induces negative emotions such as "confusion" if it is not reduced by the successful recognition of the second category. The extent of the free energy increase in controlled process $\Delta F_{ij}$ where one tries to switch one's recognition from category $i$ to the second category $j$ is calculated as the difference between the two KL divergences using Eq. 15. With the Gaussian Bayes model, the free energy increase forms a quadratic polynomial of the distance between prior means, $\mu_{ij}$, and the prediction error of category $j$ is recognized in the controlled process, $\delta_j$. The coefficients are functions of variances.

$$\Delta F_{ij} = D_{KL}[p_i(x|y)||p_j(x)] - D_{KL}[p_i(x|y)||p_i(x)]$$

$$= \frac{1}{2}\left\{\frac{s_L^2 - s_i s_j}{s_j(s_i + s_L)^2}\mu_{ij}^2 + \frac{s_i(s_i - s_j)}{s_j(s_i + s_L)^2}\delta_j^2 + \frac{2s_i(s_j + s_L)}{s_j(s_i + s_L)^2}\mu_{ij}\delta_j + \frac{s_L(s_i - s_j)}{s_j(s_i + s_L)} + \log\frac{s_j}{s_i}\right\} \quad (32)$$

With the partial differential of $\mu_{ij}$, we find that the distance between priors' means increases the free energy increase when $s_L^2 - s_i s_j \geq 0$ in the first term of Eq. 33.

$$\frac{\partial \Delta F_{ij}}{\partial \mu_{ij}} = \frac{s_L^2 - s_i s_j}{s_j(s_i + s_L)^2}\mu_{ij} + \frac{s_i(s_j + s_L)}{s_j(s_i + s_L)^2}\delta_j \quad (33)$$

When $s_L^2 - s_i s_j < 0$, the $\Delta F_{ij}$ forms a concave function of $\mu_{ij}$ with a peak at a certain distance between priors' means as a function of variances, $\mu_{ij} = \frac{s_i s_j + s_i s_L}{s_i s_j - s_L^2}\delta_j$.

By introducing the relative prediction error $r_j = \delta_j/\mu_{ij}$ $(0 < r_j < 1)$ and prior variance ratio, $s_r = s_j/s_i$, the free energy increase $\Delta F_{ij}$ is formed as a quadratic function of the distance between priors, $\mu_{ij}^2$.

$$\Delta F_{ij} = A\mu_{ij}^2 + B \quad (34)$$

$$A = \frac{1}{2s_r(s_i + s_L)^2}\left\{s_i(1 - s_r)r_j^2 + 2(s_r s_i + s_L)r_j + \frac{s_L^2 - s_r s_i^2}{s_i}\right\}$$

$$B = \frac{1}{2}\left\{\frac{s_L(1-s_r)}{s_r(s_i+s_L)} + \log s_r\right\} = \frac{1}{2}\left\{\frac{s_L}{(s_i+s_L)}\left(\frac{1}{s_r}-1\right) + \log s_r\right\}$$

The gradient $A$ of Eq. 34 is a function of four parameters: relative prediction error $r_j$, variance parameters prior variance $s_i$, $s_r$, and likelihood variance $s_L$. The intercept $B$ depends only on the variance parameters $s_i$, $s_r$, and $s_L$.

Figure 7 shows the free energy increase as functions of relative prediction error $r_j$ for different prior variances at different levels of distances between priors. We find that the relative prediction error always increases the free energy increase regardless of variance conditions by the partial differential in Eq. 35.

$$\frac{\partial \Delta F_{ij}}{\partial r_j} = \frac{\mu_{ij}^2}{(s_i+s_L)^2}\left\{\frac{s_i}{s_r}r_j + 2\frac{s_L}{s_r} + (2-r_j)s_i\right\} > 0, (\because 0 < r_j < 1) \qquad (35)$$

The prior variance ratio decreases the effect of the relative prediction error on the increase in the free energy decreases as the partial differential in Eq. 36.

$$\frac{\partial}{\partial s_r}\frac{\partial \Delta F_{ij}}{\partial r_j} = \frac{\mu_{ij}^2}{(s_i+s_L)^2}\left\{-\frac{s_i}{s_r^2}r_j - 2\frac{s_L}{s_r^2}\right\} < 0 \qquad (36)$$

The intercept $B$ of Eq. 34 is greater than zero when $\frac{1}{s_i/s_L+1} > \frac{\log s_r}{\frac{1}{s_r}-1}$. Thus, when the ratio of likelihood variance $s_L$ over the prior variance is greater than the ratio of the prior variances, the intercept tends to be positive. In conclusion, the free energy increase $\Delta F_{ij}$ tends to be positive when the relative prediction error $r_j$ is greater than a certain value and/or the likelihood variance is greater than prior variances.

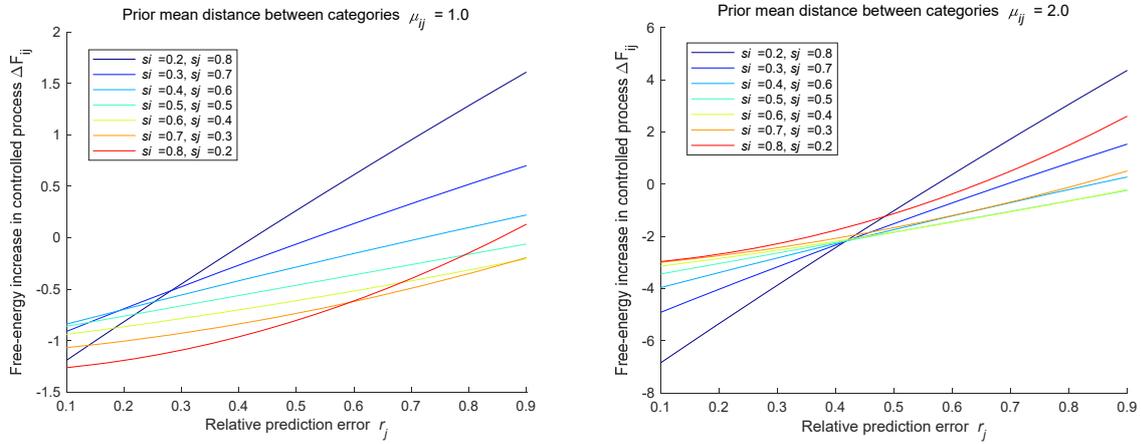

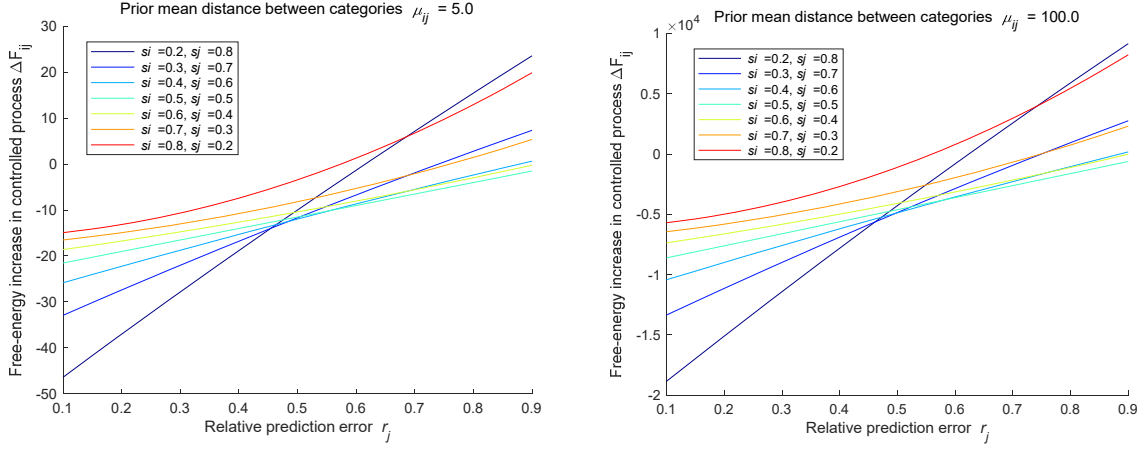

Fig. 7. Free energy increase in the controlled process (disfluency) as a function of the relative prediction error in the same process. The relative prediction error always increases the free energy increase regardless of the conditions of prior variances and distance between priors.

### 3.4 Disfluency reduction and interest: Free energy reduction in the controlled process

We associate the free energy reduction in a controlled process with disfluency reduction that induces positive emotions such as "interest." The extent of the free energy reduction with the Gaussian Bayes model is formed as a quadratic polynomial function of the distance between prior means, $\mu_{ij}$, and the prediction error of the category $j$ recognized in the controlled process, $\delta_j$. The coefficients are functions of variances:

$$\Delta F_j = D_{KL}[p_i(x|y)||p_j(x|y)] = \frac{1}{2}\left[\frac{s_L\{(s_j + s_L)\mu_{ij} + (s_i - s_j)\delta_j\}^2}{s_j(s_j + s_L)(s_i + s_L)^2} + \frac{s_i(s_j - s_L)}{s_j(s_i + s_L)} + \log\frac{s_j(s_i + s_L)}{s_i(s_j + s_L)}\right] \geq 0 \quad (37)$$

By introducing the relative prediction error $r_j = \delta_j/\mu_{ij}$ $(0 < r_j < 1)$, $\Delta F_j$ is formed as a quadratic function of the distance between prior means $\mu_{ij}^2$.

$$\Delta F_j = \frac{1}{2}\left(A\mu_{ij}^2 + B\right) \quad (38)$$

$$A = \frac{s_L\{(s_i - s_j)r_j + s_L + s_j\}^2}{s_j(s_j + s_L)(s_i + s_L)^2}, B = \frac{s_i(s_j - s_L)}{s_j(s_i + s_L)} + \log\frac{s_j(s_i + s_L)}{s_i(s_j + s_L)}$$

The gradient $A$ is a function of the relative prediction error $r_j$ and variances, whereas the intercept is a function of variances only. We find that the sign of gradient $A$ depends on the magnitude relations of priors' variances as shown in Eq. 40 by the partial differential of Eq. 39.

$$\frac{\partial \Delta F_j}{\partial r_j} = \frac{2s_L(s_i - s_j)(s_i r_j + s_L + (1 - r_j)s_j)}{s_j(s_j + s_L)(s_i + s_L)^2}\mu_{ij}^2 \quad (39)$$

$$\frac{\partial \Delta F_j}{\partial r_j} > 0, \quad if \ s_i > s_j$$

$$\frac{\partial \Delta F_j}{\partial r_j} < 0, \quad if \ s_i < s_j \quad (40)$$

$$\frac{\partial \Delta F_j}{\partial r_j} = 0, \quad if \ s_i = s_j$$

Figure 8 shows the free energy reduction as a function of the relative prediction error for different combinations of priors' variances. When the prior variance of category $i$ is greater than that of category $j$, $s_i > s_j$, the free energy reduction increases as the relative prediction error increases. By contrast, when $s_i < s_j$, the free energy reduction increases as the relative prediction error increases. When $s_i = s_j$, the free energy reduction does not change over the relative prediction error.

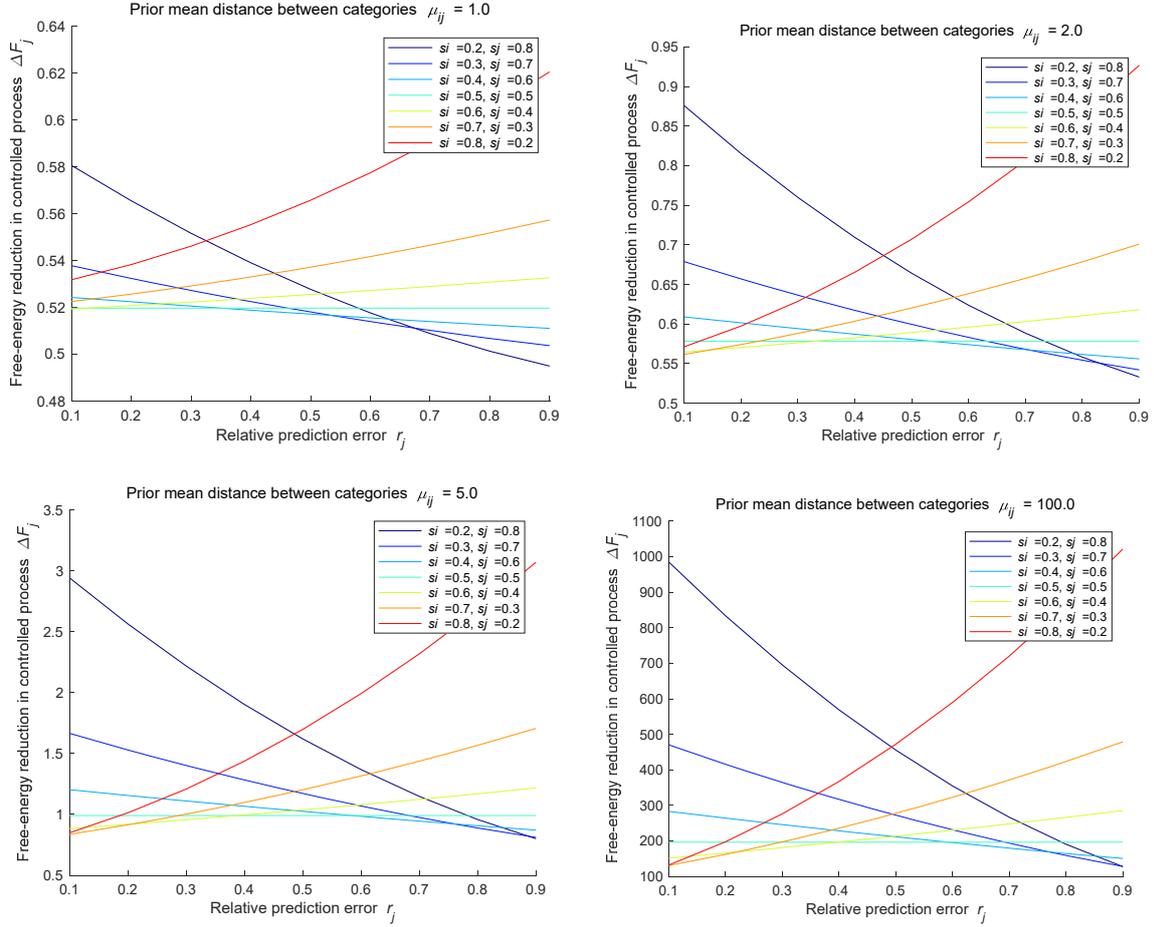

Fig. 8. Free energy reduction in the controlled process (disfluency reduction) as a function of the relative prediction error of category $j$ recognized in the same process for different priors' variance combinations. Regardless of the distance between priors' means, the free energy reduction increases when $s_i > s_j$, and decreases when $s_i < s_j$ as the relative prediction error increases. The free energy reduction is not altered by the relative prediction error when $s_i = s_j$.

We find that the distances between the priors' means always increase the free energy reduction by partial differential Eq. 41.

$$\frac{\partial \Delta F_i}{\partial \mu_{ij}} = \frac{s_L \{s_L + s_i r_j + (1 - r_j) s_j\}^2}{s_j (s_j + s_L)(s_i + s_L)^2} \mu_{ij} > 0 \qquad (41)$$

Figure 9 shows the free energy reduction as a function of the distance between priors' means for different priors' variance combinations in different levels of the relative prediction errors. The gradient of the function shown in Eq. 41 depends on $r_j$. As $r_j$ increases, the ratio of prior $s_r$ tends to decrease the gradient.

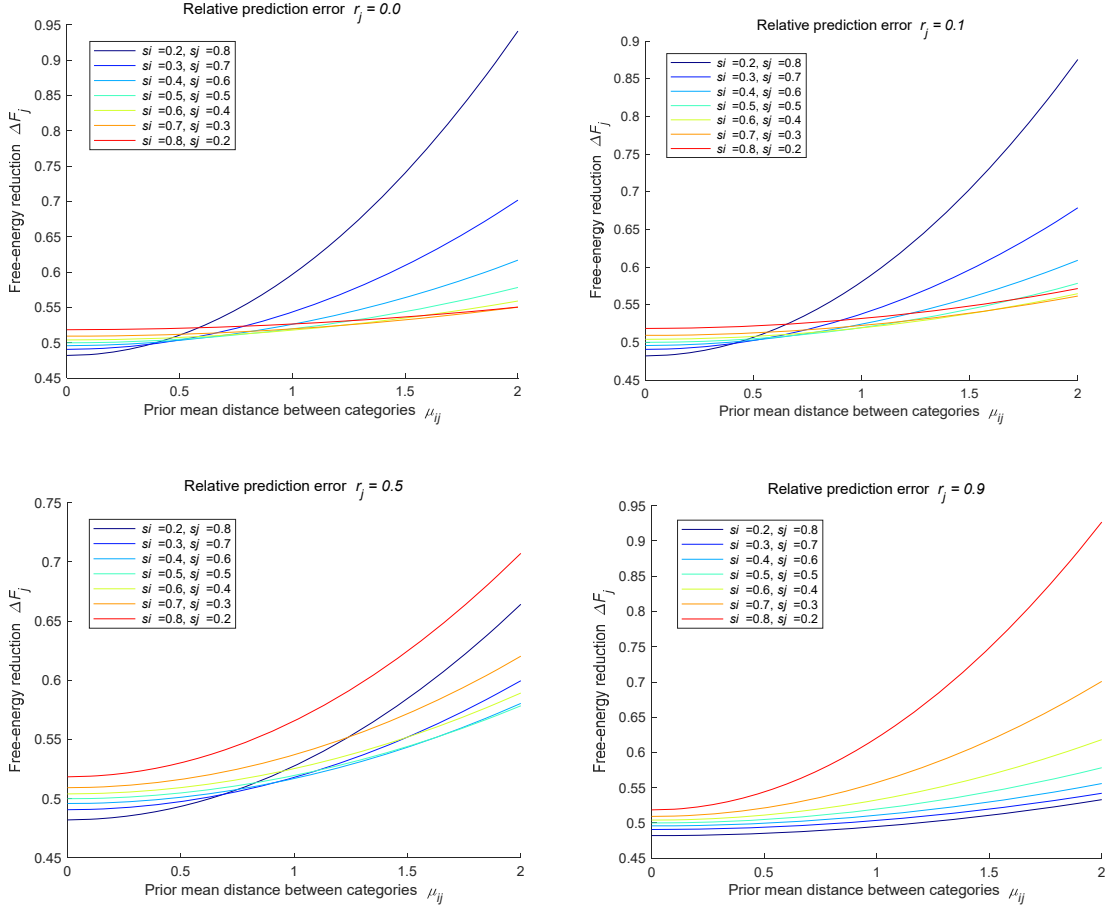

Fig. 9. Free energy reduction in the controlled process (disfluency reduction) as a function of the distance between priors' means for different priors' variance combination in three relative prediction error conditions. The free energy reduction increases with the distance between the priors' means regardless of any conditions.

## 4. Discussion

We modeled the dual process as a transition of category recognition from automatic to controlled process using a hierarchical Bayesian perception model. The free energy flow in the transition forms free energy variations (i.e., increase and decrease). In Section 2.7, we associated the free energy variations with the PIA model (Graf & Landwehr, 2015) and defined emotions such as "interest," "confusion," and "boredom" in the controlled process using the combination of two free energy variations: the free energy increase when one tries to recognize the second category $\Delta F_{ij}$ as disfluency, and the free energy reduction when one succeeds to recognize the second category $j$ in controlled process $\Delta F_j$ as disfluency reduction. We considered that "interest" involves both free energy increase and its reduction, whereas "confusion" lacks free energy reduction. "Boredom" does not involve the free energy increase, and thus no free energy reduction (Fig. 3). From Eq. 19, the free energy reduction $\Delta F_j$ consists of the summation of the free energy increase $\Delta F_{ij}$ and the difference in minimized free energy between categories, $\Delta F_{ij,min}$. In this context, $\Delta F_{ij}$ and $\Delta F_{ij,min}$ represent a reward for the effort of the controlled process (i.e., one tries to recognize the same observation as a different category.) and an absolute reduction of free energy (or information gain), respectively. Therefore, the free energy increase $\Delta F_{ij}$ is a factor of both negative and positive emotions such as confusion and interest.

We applied the Gaussian Bayesian model and found that these free energy differences form quadratic polynomials of the distance between prior means $\mu_{ij}$ and the prediction error of the second category $j$, $\delta_j$, with coefficients as functions of prior variances $s_i$, $s_j$ and the likelihood width $s_L$ as formulated in Eqs. 32 and 37. The distance between prior $\mu_{ij}$ denotes the distance in a feature dimension between the two categories. The prediction error $\delta_j$ represents the distance between the most probable (typical) feature of the second category and likelihood given by observations. The greater the prediction error, the more difficult it is to recognize the second category in the controlled process. In this sense, the prediction error $\delta_j$ denotes the difficulty in the recognition of the second category in the controlled process. The prior variance of a category represents variance or uncertainty of feature of the category belief. The prior variance is affected by both the feature distribution of the category's members and one's knowledge about the category. If the feature of the category's member varies, the prior variance becomes large. If ones' knowledge about the feature of the category is uncertain, the prior variance can be large. Table 1 summarizes the mathematical analysis using the Gaussian model on how these parameters affect free energy differences. We discuss each difference in the following sections.

Table 1. Summary of Gaussian analysis of three types of free energy variations affected by distance between priors and relative prediction error.

| | Difference in minimized free energy: $\Delta F_{ij,min} = F_{i,min} - F_{j,min}$ | Free energy increase (Disfluency) $\Delta F_{ij} = F_j - F_{i,min}$ | Free energy reduction (Disfluency reduction or Interest) $\Delta F_j = \Delta F_{ij} + \Delta F_{ij,min}$ |
|---|---|---|---|
| Relative prediction error $r_j = \dfrac{\delta_j}{\mu_{ij}}$ | Always decrease (Fig. 6) <br> Concave $(s_i < s_j)$ <br> Convex $(s_i > s_j)$ <br> Linear $(s_i = s_j)$ | Always increase. (Fig. 7) | Increase $(s_i > s_j)$ <br> Decrease $(s_i < s_j)$ <br> Constant $(s_i = s_j)$ <br> (Fig. 8) |
| Distance between priors' mean $\mu_{ij}$ | Increase, $\dfrac{(1-r_j)^2}{s_i+s_L} > \dfrac{r_j^2}{s_j+s_L}$ <br><br> Decrease, $\dfrac{(1-r_j)^2}{s_i+s_L} < \dfrac{r_j^2}{s_j+s_L}$ <br><br> Constant, $\dfrac{(1-r_j)^2}{s_i+s_L} = \dfrac{r_j^2}{s_j+s_L}$ | Increase if $s_L^2 > s_i s_j$; otherwise, concave with a peak of $\mu_{ij} = \dfrac{s_j+s_L}{1-\frac{s_L^2}{s_i}}$. | Always increase. (Fig. 9) |

### 4.1 Likelihood of category recognitions in the dual process

The difference in minimized free energy, $\Delta F_{ij,min}$, denotes a relative likelihood to be recognized as the second category $j$ than the initial category $i$ ($F_{i,min} > F_{j,min}$). This difference is a factor to increase free energy reduction. The greater the relative likelihood, the more convincing the recognition in the controlled process. The sign of the free energy difference determines which category is more likely to be recognized.

The Gaussian analysis shows that the relative prediction error always decreases the relative likelihood of the category (Fig. 6). The distance between priors $\mu_{ij}$ decreases the relative likelihood when the relative prediction error weighted by the precisions (inverse variances) of the second category $j$ is greater than that of category $i$, i.e., $r_j^2/(s_j + s_L) > (1 - r_j)^2/(s_i + s_L)$ from Eq. 31. Under this condition, the relative likelihood is always negative when the prior variance of the controlled process is greater than that of the automatic process, $s_j > s_i$. By contrast, the relative likelihood is always positive if the relative prediction error weighted by the precisions of the second category $j$ is smaller than that of category $i$, $r_j^2/(s_j + s_L) < (1 - r_j)^2/(s_i + s_L)$, and prior variance of the controlled process is smaller than that of automatic process, $s_j < s_i$. Thus, the prior variances affect the relative likelihood of the recognized category, as well as prediction error.

When $\Delta F_{ij,min}$ is negative (i.e., $F_{i,min} < F_{j,min}$), the initial category recognized in the automatic process is likely to be recognized than the second category recognized in the controlled process. This condition is likely because the controlled process needs the effort to recognize the second category, whereas the automatic process is effortless in recognizing the initial category. In the WML case, the young lady may be more likely recognized than an old lady, and the likelihood of recognizing a young lady is greater than that of an old lady even after succeeding in the recognition of an old lady as the second category in the controlled process.

However, the inverse case, $\Delta F_{ij,min}$ is positive (i.e., $F_{i,min} > F_{j,min}$), can occur when one cannot initially find a specific category to fit observations due to noisy or complex stimuli. In this case, the initial prior is flat, and the free energy reduction is small; hence, the minimized free energy remains in a high state. For example, in Gregory's Hidden Dalmation Dog (Gregory, 2005), one initially sees the random black and white dots and then finds the hidden Dalmation dog in the pattern of the dots. After succeeding in the recognition, one recognizes the Dalmation dog more than random dots. In this case, the second category (Dalmation dog) is more likely to be recognized than the initial recognition (random dots), thus $\Delta F_{ij,min}$ is positive.

### 4.2 Characteristics of emotions in the controlled process

Emotions in the controlled process are determined by the free energy increase $\Delta F_{ij}$ and its reduction $\Delta F_j$, representing disfluency and disfluency reduction, respectively. The Gaussian analysis suggests that the relative prediction error of the second category, $r_j$, always increases the free energy increase (i.e., disfluency) as shown in Fig. 7. However, the relative prediction error both increase and decrease the free energy reduction (disfluency reduction) depending on the magnitude relation of prior variances, as shown in Fig. 8. When the prior variance of the controlled process is smaller than that of the automatic process, $s_j < s_i$, the relative prediction error increases the free energy reduction. By contrast, when $s_j > s_i$, the relative prediction error decreases the free energy reduction. When $s_j = s_i$ the relative prediction error does not affect the free energy reduction.

Free energy denotes uncertainty and prediction error as discussed using Eqs. 23 and 24, and its reduction induces positive emotions (Joffily & Coricelli, 2013). We consider that the resolution of greater uncertainty and prediction error induces more positive emotions. In this context, the Gaussian analysis suggests that the greater (relative) prediction error induces more positive emotions by succeeding in the recognition of the second category when the prior variance of the controlled process is smaller than that of the automatic process. By contrast, the smaller prediction error induces more positive emotions when the prior variance of the controlled process is greater than that of the automatic process. The relative prediction error does not affect emotions when the prior variances of the two processes are the same.

This dependency based on the magnitude relations of prior variances is due to the characteristic of KL-divergence shown in Eq. 18 (i.e., the definition of free energy reduction or disfluency reduction), not the Gaussian distribution. The prior's variance increases the posterior's variance. The prediction error of category $j$ increases the distance between the posterior's means. The distance increases the KL-divergence of Eq. 18 when the posterior probabilities of category $i$ are distributed over feature values where the overlap of the two posteriors is small. This condition is realized when the posterior variance of category $i$ is greater than that of variance $j$ such that the posterior distributions of $i$ cover the less overlap part. This suggests that the dependency based on priors' variance is applied to any type of distribution as well as Gaussian. From Eq. 19, the free energy reduction $\Delta F_i$ is a summation of two terms: free energy increase $\Delta F_{ij}$ and the difference in minimized free energy between categories, $\Delta F_{ij,min}$. As Table 1 shows, the prediction error increases $\Delta F_{ij}$ and decreases $\Delta F_{ij,min}$. Therefore, the positive effect of the prediction error on free energy reduction is due to the free energy increase meaning disfluency.

Both disfluency and its reduction are required to induce positive emotions such as "interest." The lack of disfluency and its reduction may lead to "boredom." Disfluency without its reduction induces negative emotions such as "confusion." Therefore, the proposed framework suggests that one experiences "interest" by recognizing the second category when the category involves a greater prediction error and a smaller prior variance (uncertainty) than the initial category.

The distance between prior means $\mu_{ij}$ always increases the free energy reduction (disfluency reduction) as shown in Fig. 9. This result suggests that the farther the initial category is from the second category in a feature dimension, the more interest one will experience. However, the distance between priors can decrease the free energy increase (disfluency) when the likelihood width is smaller than priors' variances, $s_L^2 < s_i s_j$, and the relative prediction error is greater than the certain value comprising the variances, $(s_j + s_L)/(1 - s_L^2/s_i)$. These results suggest that both priors' distance and prediction error may suppress the disfluency when prior variances are larger than the likelihood variance. The free energy increase (disfluency) is always positive if $\frac{1}{s_i/s_L+1} > \frac{\log s_r}{1/s_r-1}$ and the above condition where the distance between prior means $\mu_{ij}$ increases the free energy increase. The former condition tends to be true when the prior variance $s_i$ is smaller than the likelihood width and the relative prior variance $s_r = s_j/s_i$ is large.

### 4.3 Motivation for the controlled process and emotions

The feeling of disfluency is uncomfortable owing to increasing free energy (uncertainty and prediction error) and may motivate one to invest more effort in subsequent conscious processing to resolve such an uncomfortable state. The PIA model suggests that the disfluency with expected fluency below its expectation in the automatic process triggers one's motivation on shifting towards the controlled process (Graf & Landwehr,

2015). By associating disfluency with free energy variations, we consider that the increase in free energy $\Delta F_{ij}$ (disfluency) with less than expected free energy reduction $\Delta F_i$ in the automatic process works as a motivation towards the controlled process. By contrast, the motivation is inhibited by satisfying pleasure in the automatic process induced by more than expected free energy reduction $\Delta F_i$ (fluency) and less increase in free energy $\Delta F_{ij}$ (disfluency). The increase in free energy $\Delta F_{ij}$ is expected to be reduced by succeeding in the controlled process. Reduction of free energy implies gaining information and resolution of uncertainty and thus induces positive emotions. This expectation may work as an incentive towards an effortful controlled process.

Disfluency $\Delta F_{ij}$ with less than expected fluency $\Delta F_i$ causes negative emotions (e.g., discomfort or confusion), but disfluency reduction causes positive emotions (e.g., interest). Therefore, the emotional valence changes from negative to positive in the transition of the dual process. Previous experiments revealed that the longer the stimuli (e.g., ambiguous paints and pictures) were presented, the more positively the participants evaluated them (Graf & Landwehr, 2017; Jakesch et al., 2013), supporting the positive effect of disfluency reduction.

From the Gaussian analysis summarized in Table 1, the relative prediction error of the second category, $r_j$, increases the fee energy increase (disfluency). The prediction error increases the free energy reduction (disfluency reduction) if a variation of the second prior is smaller than that of the initial one. These causalities suggest that the relative prediction error and uncertainty (prior variance) of the second category feature are key factors to motivate one to shift toward the controlled process. The prediction error denotes the distance between observation (likelihood) and the second category knowledge (prior distribution). The greater the prediction error, the more information processing is required. Thus, the prediction error increases the processing load of the brain, in other words, the extent of effort.

The increase in free energy is expected to be reduced by the successful recognition of the second category. This expectation of free energy reduction is increased by the prediction error when the prior brief or knowledge about a feature of the second category is certain (small prior variance). The PIA model suggests that prior knowledge is another trigger of motivation towards the controlled process where one needs cognitive enrichment. Therefore, the less prior variance due to uncertainty of knowledge about the second category may motivate one to challenge more effortful cognitive processing that is expected to provide greater positive emotions when resolved (i.e., interest!). This free energy reduction may represent "Aha moment"/"perceptual insight" and its positive emotion (Chetverikov & Filippova, 2014; Muth et al., 2013; Muth, Raab, et al., 2015).

### 4.4 Expected free energy reduction and information processing capacity

The expectation of free energy reduction means information gain is expected and may trigger curiosity, as discussed in a previous study (Friston et al., 2017). This expectation is realized as the decreasing absolute state of free energy $\Delta F_{ij,min}$ in our model. In addition to the decrease, our model suggests that the prior switching for representing the transition of the dual process increases the diminution of free energy $\Delta F_j$ by its temporal increase $\Delta F_{ij}$ as discussed using Eq. 19. This temporal increase in free energy with the expectation of its reduction may motivate one to invest cognitive effort. In our model, this motivation is triggered by the (expected) free energy drop from the increased free energy state $F_j$, and that diminution $\Delta F_j$ induces a positive emotion. As Eq. 18 shows, this free energy reduction always occurs after subsequent recognitions of the second category, even if the absolute state of the free energy increases, i.e., $\Delta F_{ij,min} = \Delta F_{i,min} - \Delta F_{j,min} < 0$.

The brain has the capacity to process information or cognitive load. The free energy state or surprise is interpreted as contents to be processed by the brain. It represents the arousal potential induced by information due to factors such as complexity and novelty (Yanagisawa, 2021). Therefore, the increase in free energy $\Delta F_{ij}$ needs to be less than a certain capacity to motivate towards an effortful controlled process. If the free energy increase exceeds the capacity, one may give up on trying the controlled process. This characteristic may explain Berlyne's theory of arousal potential, suggesting that an appropriate level of arousal potential induces a positive hedonic response and motivation to approach to the stimuli, whereas extreme arousal potential induces negative response and avoidance (Berlyne, 1970). The inverse U shaped hedonic function of arousal potential (Wudnt curve) is formed by the summation of two sigmoidal functions: reward and aversion (Yanagisawa et al., 2019). Arousal potential increases both rewards (positive response) and aversion (negative response). The increase in free energy and its (expected) reduction may represent the reward function, whereas the capability to process free energy increase may explain the initial rise in the arousal potential of the aversion system.

In addition to capacity, personality and motivation are also important to discuss free energy reduction. Several experimental studies have reported that the appreciation of recognizing an ambiguous artwork is positively and negatively related to the difficulty in recognition, depending on the personality (ambiguity tolerance) and motivational mode (preservation/promotion mode) affected by the experimental situation (Muth, Hesslinger, et al., 2015; Muth, Raab, et al., 2015).

The neural basis for the transition from automatic to controlled process is unclear. Automatic processing to monitor the external environment (i.e. rapid processing for a particular predictable context) is realized by default mode brain networks in the dual process of thinking (Buckner et al., 2008; Vatansever et al., 2017). In a series of studies on the predictive and reactive control systems (PARCS) model by Tops and colleagues, it is shown that the cognitive resources available for monitoring the environment are limited and that the emotions and motivations generated depend on the need for cognitive effort and the specific context (Tops, 2017; Tops et al., 2017). It has also been reported that brain processing differs, including the default mode brain network involved in these processes.

### 4.5 Potential experiments linking the proposed framework to neural basis

To experimentally verify whether the mathematical model and analysis proposed in this study are realized in the human brain, it is necessary to measure neural activity during transitions in dual-process perception using functional brain imaging techniques. Many neuroscience studies related to dual processes use functional magnetic resonance imaging (fMRI) (Kim et al., 2007; Vatansever et al., 2017). Brain regions involved in emotions (Kim et al., 2007) and flexibility (Vatansever et al., 2017) in the automatic and controlled processes can be verified with detailed location information by fMRI with high spatial resolution. By contrast, especially for automatic processing, which is rapid, the detailed understanding of temporal dynamics requires investigations with electroencephalograms (Tops et al., 2017) and event-related brain potentials (Mansfield et al., 2013), which have a superior temporal resolution.

It is also important to determine the experimental paradigm that should be used to examine the transition from automatic to controlled processes in the dual-process perception. Priming tasks, such as having participants make face preference judgments (Kim et al., 2007), answer a quiz (Ligneul et al., 2018) twice with a time delay, or the Wisconsin Card Sorting Test (Vatansever et al., 2017) to assess cognitive sets, allows for the experimental manipulation of parameters such as pre-and post-event uncertainty and distance between categories in situations where categories shift.

### 4.6 Limitations and further discussions

In Section 3, we used a Gaussian Bayes model of single dimension variables for convenience to analyze the basic characteristics of the effect of expectations (i.e., the distance between priors' means and prediction error of the second category recognized in the controlled process) and prior variations (i.e., uncertainty and inverse precision) on emotions in the dual process using free energy variations. The real problems may, however, follow different complex distributions with multiple variables. Nevertheless, the framework proposed in Section 2 is a general form, and thus it can be applied to any type of distribution of multiple variables. Analysis using different distributions with multiple variables in a variety of specific problems may elucidate further potential characteristics of emotions in the dual process.

Observation $y$ in the model may change owing to the observer's actions and stimulus changes. In such dynamic observation settings, the free energy may fluctuate during the controlled process. However, free energy reduction (disfluency reduction) is assumed to have instantaneously occurred as discussed in Section 2. Thus, the observations made immediately before and after the recognition are assumed to be equivalent. Therefore, the analysis and discussion of the free energy reduction using the same observation is valid. Yet, the free energy increase (i.e., disfluency) may fluctuate until successfully recognizing the second category by changing the observations. Further discussion is expected for modeling dynamical cognition during the controlled process using the free energy variations.

Generative models of categories $p(x,y|c)$ may involve individual differences due to individual knowledge, prior experience, personality/mode, and so forth. The difference between the generative models fluctuates the free energy (reduction), causing differences in experimental results between participants. In addition, individual personality may affect the motivation/capacity to cope with the free energy increase (toward controlled process). The framework is expected to explain such individual difference effects by using the uncertainty of prior as prior knowledge and experience (Miyamoto & Yanagisawa, 2021) and acceptance of arousal potential (i.e., free energy increase) as a threshold of positive response to the effort of the controlled process (Yanagisawa et al., 2019).

### 5. Conclusion

This paper presented a proposed general mathematical framework of emotional valence in dual-process perception based on free energy dynamics in a hierarchical Bayesian model. The model formulates emotions such as pleasure, interest, confusion, and boredom using free energy variations in a shift from the automatic to the controlled process represented by Bayesian prior switching. The model mathematically demonstrates that the

"interest" as a positive emotion in the controlled process requires the overcoming of the effortful process represented as free energy increase, unlike "pleasure," as another positive emotion in the automatic process. This explains the motivation to challenge effortful tasks such as trying to cope with complex and novel stimuli or events, rather than staying in comfortable effortless states. The Gaussian model-based analysis exhaustively reveals the effects of prediction errors and uncertainties on the extent of "interest." These results provide general and fundamental knowledge to increase such positive emotions and motivations to challenge effortful tasks because the proposed model is deduced from the synthesis of free energy minimization as the first principle of a biological system (e.g., the brain) and the well-established fluency–disfluency paradigm in dual-process perception. Therefore, this model can be applied to diverse areas dealing with emotion and motivation, such as education, creativity, aesthetics, affective computing, and related cognitive sciences. Further studies in both theoretical and experimental approaches are desired for understanding such emotions in more complex situations and applying the principle-based model to diverse disciplines.

## Declaration of Competing Interest

The authors declare that they have no known competing financial interests or personal relationships that could have appeared to influence the work reported in this paper.

## Funding

This research was supported by the Japan Society for the Promotion of Science (KAKENHI Grant Number 21H03528, Mathematical model development of emotion dimensions based on variation of uncertainty and its application to inverse problems).